# Two distinct notions of speed in antiferromagnets

**Paul Herrgen,[1*] Stephan Wust,[1] Christin Schmitt,[2] Yaryna Lytvynenko,[2] Lorenzo Baldrati,[2] Rafael Ramos,[3,4] Takashi Kikkawa,[5,6] Eiji Saitoh,[4,5,7,8] Martin Jourdan,[2] Mathias Kläui,[2] Hans Christian Schneider, [1] Benjamin Stadtmüller,[1, 9] and Martin Aeschlimann[1]**

[1]*Department of Physics and Research Center OPTIMAS, RPTU University Kaiserslautern-Landau, 67663 Kaiserslautern, Germany*

[2]*Institute of Physics, Johannes Gutenberg University Mainz, 55099 Mainz, Germany*

[3]*CIQUS, Departamento de Química-Física, Universidade de Santiago de Compostela, Santiago de Compostela, Spain*

[4]*WPI Advanced Institute for Materials Research, Tohoku University, Sendai 980-8577, Japan*

[5]*Department of Applied Physics, The University of Tokyo, Tokyo 113-8656, Japan*

[6]*Advanced Science Research Center, Japan Atomic Energy Agency, Tokai 319-1195, Japan*

[7]*Institute for AI and Beyond, The University of Tokyo, Tokyo 113-8656, Japan*

[8]*RIKEN Center for Emergent Matter Science (CEMS), Wako 351–0198, Japan*

[9]*Institute of Physics, University of Augsburg, 86159 Augsburg, Germany*

**Corresponding author. E-mail: pherrgen@rptu.de*

## Abstract

The terahertz resonance frequencies of antiferromagnets characterize their precessional dynamics but do not necessarily imply faster laser-induced quenching of magnetic order. Here, we investigate the distinct roles of magnetic compensation and electronic scattering phase space by comparing ultrafast demagnetization in the antiferromagnetic charge-transfer insulator NiO, the metallic antiferromagnet $Mn_2Au$, and a ferromagnetic Ni reference. The extracted response in $Mn_2Au$ is approximately four times faster than in NiO and substantially faster than in Ni. These observations are consistent with exchange-mediated scattering between antiparallel sublattices enabling rapid loss of antiferromagnetic order without requiring net angular-momentum transfer to an external reservoir, provided sufficient electronic scattering phase space is available. In insulating NiO, restricted electronic scattering channels reduces the ultrafast response. Our results distinguish ultrafast demagnetization from exchange-enhanced precession and identify the interplay of spin-flip probability and electronic scattering rate as a key factor governing the nonequilibrium response of compensated magnets.

## Introduction

Antiferromagnets (AFMs) are frequently characterized as "intrinsically fast" due to their terahertz (THz) resonance frequencies[1-3], which exceed the gigahertz (GHz) scales typical of ferromagnets[4] (FMs). This speed refers specifically to precessional dynamics in quasi-equilibrium, governed by the Landau-Lifshitz-Gilbert (LLG) framework.

$$\frac{d\boldsymbol{m}}{dt} = -\gamma\mu_0 \boldsymbol{m} \times \boldsymbol{H}_{eff} + \alpha \boldsymbol{m} \times \frac{d\boldsymbol{m}}{dt} \quad (1)$$

For a simple, easy-axis, two-sublattice antiferromagnet, a representative resonance frequency for AFMs is

$$\omega_{AFM} \sim \gamma\mu_0 \sqrt{H_{ex} H_{an}} \quad (2)$$

where $H_{ex}$ is the exchange field and $H_{an}$ is the effective anisotropy field. Together, the large exchange field and the anisotropy field produce a high resonance frequency in AFMs.

In contrast, ultrafast laser-induced demagnetization - the quenching of the magnetic order parameter on a timescale $\tau_m$ (hereinafter referred to as the demagnetization time $\tau_m$, also for AFMs) - is a far-from-equilibrium process where these fields $H_{ex}$ and $H_{an}$ play a negligible role. Consequently, the widely cited THz speed of AFMs does not fundamentally imply a shorter demagnetization time $\tau_m$ or enhanced damping. Thus, there exist two notions of "speed" that are physically distinct, but very little is known about their relationship.

In FMs, ultrafast demagnetization is typically described by electronic scattering processes, in which the electronic spin state is changed due to electronic quasi-particle interactions including spin-orbit coupling (see Fig. 1a). These processes can be interpreted in a framework originally developed for spin relaxation processes and by Elliott and Yafet. In this picture, the demagnetization rate is governed by the product of the scattering rate and the spin-flip probability $a_{sf}$[5-7], the latter being a direct consequence of spin-mixing in the electronic states. A critical bottleneck in FMs is the requirement to dissipate angular momentum into an external reservoir, such as the lattice, to facilitate this quench.

Collinear compensated AFMs, offer a distinct advantage: because the net magnetic moment is zero, the total spin angular momentum can remain unchanged even as the AFM order parameter is destroyed. This allows for additional, highly efficient relaxation channels, specifically exchange-mediated electron-electron Coulomb scattering, which includes a spin flip in both magnetic sublattices (see Fig. 1b). Such exchange scattering processes can also be assigned a spin-flip probability, which we also denote by $a_{sf}$ even though the the microscopic interaction is different from that in ferromagnets. Since in this case the underlying exchange interaction does not need a spin-orbit coupling contribution. $a_{sf}$ can be significantly increased without necessitating angular momentum transfer to the lattice. This is a second, independent concept of “speed” in antiferromagnets. Evidence for such accelerated dynamics has already been found in 4f AFMs, where RKKY-mediated spin-spin scattering channels between localized moments can lead to faster demagnetization in the AFM phase compared to the FM phase[8]. These exchange-scattering effects also underlie the ultrafast sublattice magnetization dynamics in ferromagnets[7,9].

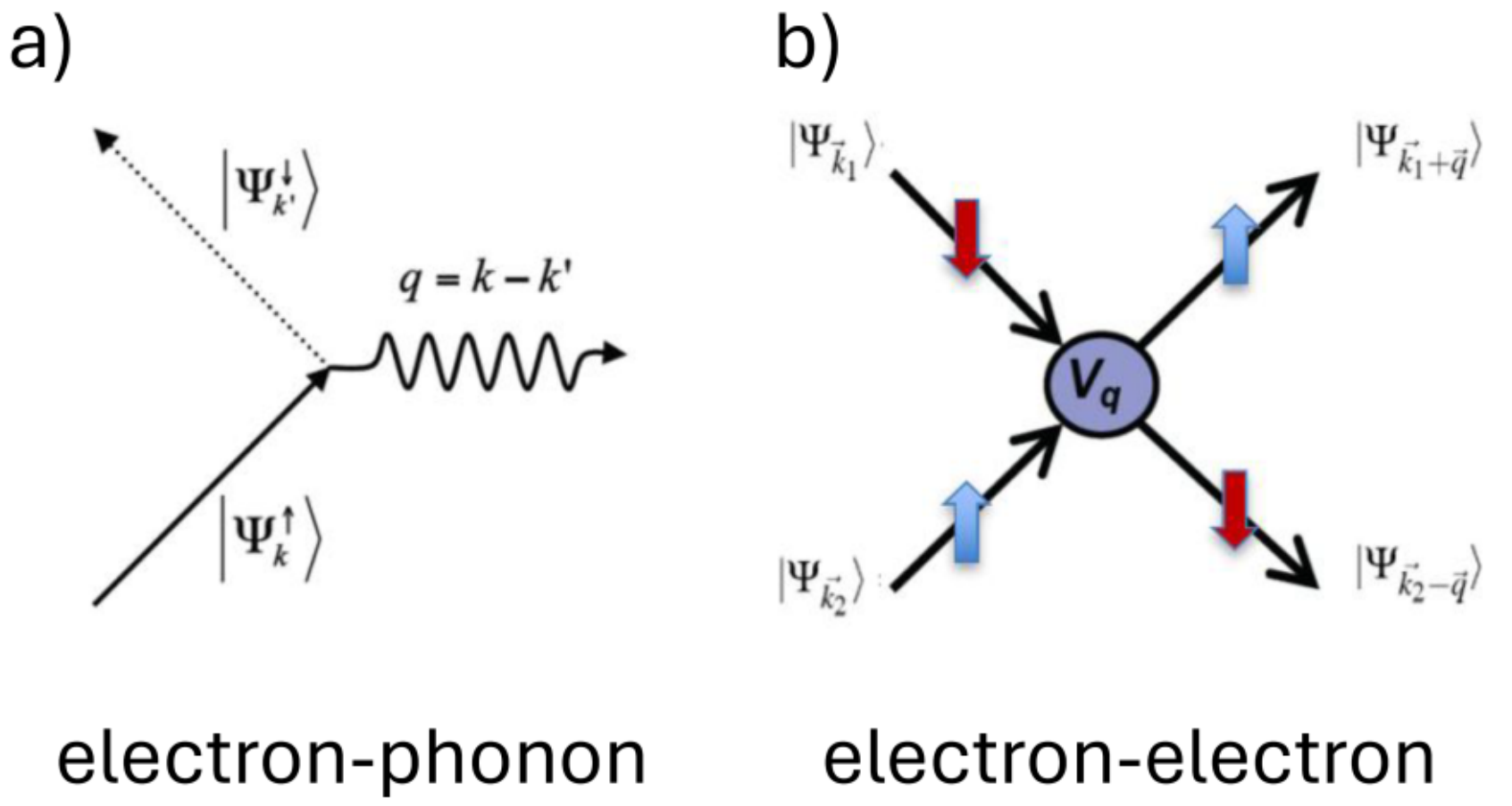

*Fig 1: Two kinds of spin-flip scattering processes. a) Elliott–Yafet type of scattering, b) Exchange-mediated electron–electron Coulomb scattering which includes a spin flip in both magnetic sublattices.*

Many prototypical AFMs are insulators introduce an effect: The presence of a band gap drastically reduces the overall scattering rate of photoexcited electrons compared to metallic systems. This creates a fundamental tension in dielectric AFM dynamics: does the increased spin-flip probability $a_{sf}$ inherent to magnetic compensated systems outweigh the suppression of scattering rates caused by the insulating gap? Furthermore, in metallic AFM does the lack of angular momentum constraints fundamentally lead to faster quenching of the order parameter (demagnetization time $\tau_m$) compared to metallic FMs?

In this Letter, we address these questions by comparing the ultrafast magnetic response of the AFM charge-transfer insulator NiO and the metallic AFM $Mn_2Au$ against a ferromagnetic Ni reference. By utilizing magneto-optical birefringence (MOB) in NiO and magnetic linear dichroism (MLD) in $Mn_2Au$, we isolate the dynamics of the AFM order parameter (demagnetization time $\tau_m$). Our results provide a critical comparison between the roles of electronic conductivity (scattering time) and angular momentum compensation, clarifying the physical limits of "speed" in antiferromagnetic spintronics.

## Experimental Methods

### Sample and excitation scheme

The NiO sample consists of a 10 nm NiO(001) film epitaxially grown on MgO(001) by reactive magnetron sputtering. The AFM order in such films induces a rhombohedral lattice distortion along the cubic [111] directions, leading to antiferromagnetic T- and S-domains[10]. For thin films, S-domains with the highest out-of-plane projections are favored, reducing it to the four T-domains only[11]. Owing to the strong magnetoelastic coupling, the magnitude of this distortion scales with the AFM order parameter and can thus be used as an optical proxy for the magnetization dynamics.

The $Mn_2Au$ sample consists of a 40 nm thick $Mn_2Au$ film grown on $Al_2O_3$(1-102) with a 13nm Ta(001) buffer stack and a 2nm $Si_3N_4$ capping by radio-frequency magnetron sputtering. $Mn_2Au$ crystallizes in a body-centered tetragonal structure with collinear AFM order, in which the Neél vector lies in the film plane[12]. Due to strong SOC from the Au 5d states, the AFM order induces an in-plane anisotropy in the electronic band structure that manifests itself as magnetic linear dichroism (MLD) in optical reflectivity.

Static measurements were performed on both samples to ensure that the measured spot was either a single-domain spot or consisted predominantly of a specific domain state. This ensures a strong measurement signal for the time-resolved measurements, since signals originating from different domains would cancel each other out.

All time-resolved measurements were performed using a regenerative amplifier providing ~35 fs pulses at a central wavelength of 800 nm with a repetition rate of 1 kHz. For NiO, the pump and probe pulses were generated by frequency tripling and doubling the fundamental to obtain photon energies of 4.65 eV (pump) and 3.10 eV (probe), thereby exciting carriers across the NiO band gap by the pump. For $Mn_2Au$, the pump photon energy was 3.10 eV (second harmonic), and the probe photon energy was 1.55 eV (fundamental), the probe energy was chosen to match a pronounced anisotropy in the band structure identified in previous photoemission experiments[12]. The absorbed pump fluence was varied over a broad range for each sample, with calibration considering reflectivity and optical penetration depth. All experiments were performed at room temperature.

### Magneto-optical birefringence in NiO

The rhombohedral lattice distortion induced by AFM leads to anisotropic reflectivity, characterized by two principal axes oriented parallel to and perpendicular with respect to the in-plane projection of the distortion. When a linearly polarized probe pulse with an electric field at an angle $\varphi$ to this projection is incident upon the sample, the polarization rotates by an angle $\phi_{pol}$, which is proportional to

$$\phi_{pol} \propto (r(\vec{T}_{\parallel}) - r(\vec{T}_{\perp}) \sin(2\varphi), \tag{3}$$

where r|| and r⊥ represents the reflectivities along the two principal axes[13]. This gives rise to a magneto-optical birefringence (MOB) effect. The magnitude of the MOB is proportional to the rhombohedral distortion, which in turn scales with the square of the sublattice magnetization. Therefore, the MOB signal can be taken as proportional to $M^2$ and can be exploited to monitor the AFM order dynamics (demagnetization time $\tau_M$) in time-resolved optical pump–probe experiments[13]. However, one should bear in mind that this is an upper bound due to possible inertia of the lattice distortion relative to the ultrafast magnetization dynamics.

To isolate the magnetic contribution from purely optical effects, we exploit the π-periodicity of the birefringence signal with respect to the sample angle. For each pump fluence, time-resolved signals are recorded at two sample orientations that yield maximum positive and maximum negative static MOB signals, corresponding to probe polarizations at ±45° with respect to the anisotropy axis, as indicated in Figure 2. Subtracting these two traces eliminates symmetric optical artifacts and yields a purely magnetic signal that can be normalized to the pre-pump value, as it has previously been done by Wust et al. for a Pt/NiO bilayer system[14]. Taking the square root of this normalized signal provides a quantity proportional to the sublattice magnetization.

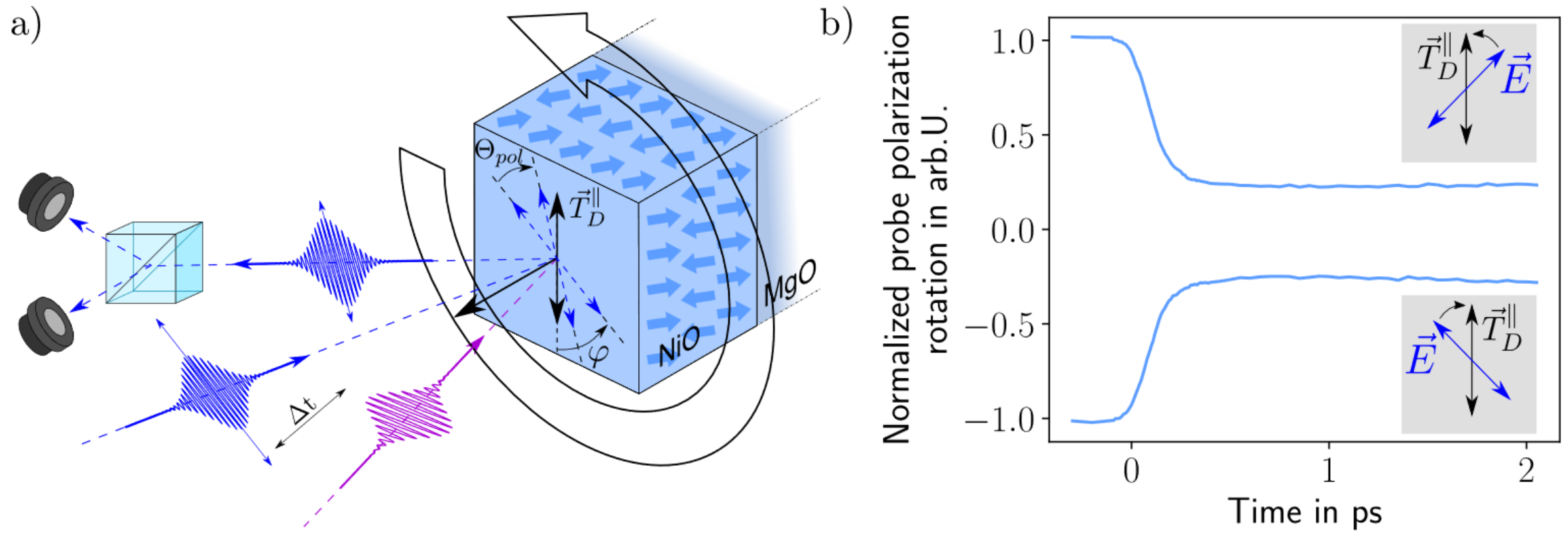

*Figure 2: a) Schematic illustration of the NiO measurement. A pump pulse of 4.65 eV (≙ 267 nm) and a probe pulse of 3.10 eV (≙ 400 nm) have been used to measure the MOB induced by the lattice distortion $\vec{T}^{\parallel}_D$. b) MOB measurements ±45° with respect to the anisotropy axis.*

## Magneto-optical birefringence in $Mn_2Au$

$Mn_2Au$ is a metallic AFM with degenerate bands and pronounced SOC originating from 5d Au states[15]. The AFM order breaks the fourfold symmetry of the in-plane band structure[12], leading to an anisotropic reflectivity for light polarized parallel and perpendicular to the Neél vector[16]. This MLD effect can be described by an expression analogous to the MOB in NiO,

$$\phi_{pol} \propto (r(\vec{N}_{\parallel}) - r(\vec{N}_{\perp}) \sin(2\varphi), \tag{4}$$

where now the axes are defined with respect to the in-plane Neél vector rather than a lattice distortion. The MLD amplitude is proportional to $M^2$ and thus provides a direct electronic probe of the AFM order parameter without involving a structural response.

The experimental geometry, shown in Figure 3 mimics that used for NiO: the sample is rotated rather than the polarization, ensuring that the relative pump–probe polarization remains fixed and that birefringence from optical components does not contaminate the signal. Two time-resolved signals at ±45° with respect to the in-plane Neél vector are recorded and subtracted to remove symmetric optical contributions, yielding a signal proportional to the time-dependent MLD. After normalization and square-root transformation, we obtain a measure of the sublattice magnetization dynamics in $Mn_2Au$.

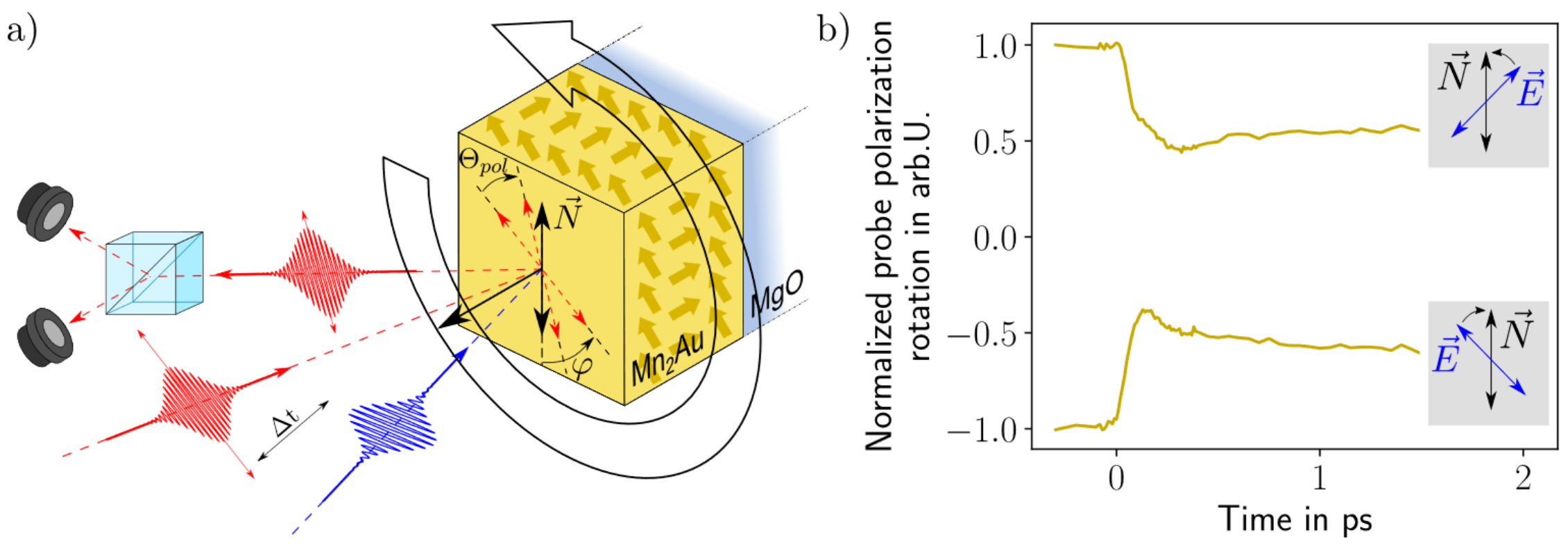

*Figure 3: a) Schematic illustration of the Mn2Au measurement. A pump pulse of 3.10 eV (≙ 400 nm) and a probe pulse of 1.55 eV (≙ 800 nm) have been used to measure the MLD induced by the Neél vector $\vec{N}$ . b) MLD measurements ±45° with respect to the anisotropy axis.*

## Data analysis

For both materials, the normalized magneto-optical signals are fitted with a double-exponential model convolved with a Gaussian temporal instrument response that accounts for the pump pulse duration. The fitting function reads

$$S(t) = [1 - Q(1 - e^{-\frac{t}{\tau_m}}) + Ae^{-\frac{t}{\tau_r}}]\, \Theta(t) + C\,, \tag{5}$$

where Q is the quenching (relative maximum reduction of the order parameter), $\tau_m$ is the demagnetization time of each sublattice, $\tau_r$ is the relaxation time associated with the recovery toward thermal equilibrium, and $\Theta(t)$ is the Heaviside function.

## Results

### a) Time-resolved MOB measurements of NiO

Figure 4a) shows the time-resolved MOB signals in NiO for absorbed pump fluences ranging from ~0.18 mJ cm$^{-2}$ to ~2.17 mJ cm$^{-2}$. All traces exhibit a rapid decrease of the AFM order parameter followed by a slower recovery toward a quasi-equilibrium state. Coherent oscillations attributable to magnons or phonons are absent within the noise level, as confirmed by Fourier analysis.

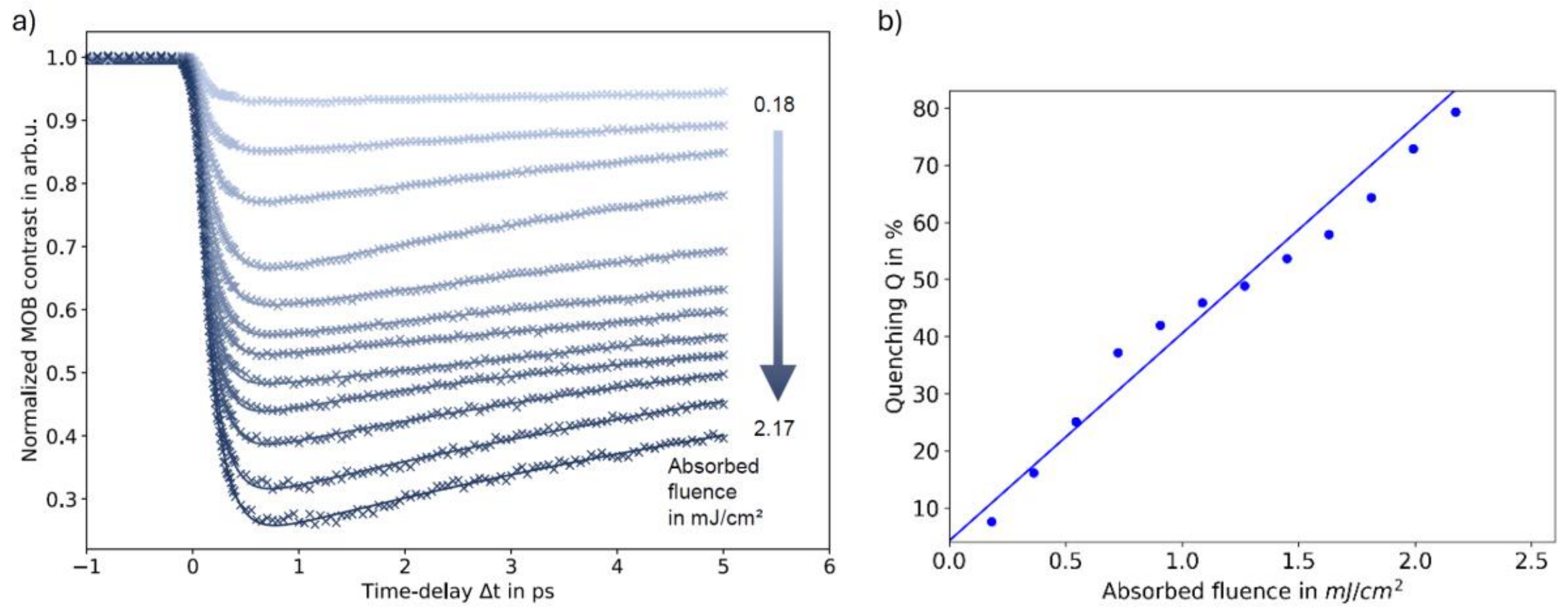

*Figure 4: a) Time-resolved MOB signal in NiO for various absorbed pump fluences. Solid lines show double-exponential fits convolved with the temporal width of the pump pulse. b) Maximum quenching Q vs. absorbed fluence*

The maximum quenching Q increases approximately linearly with absorbed fluence (Fig.4b), consistent with the behavior observed in 3d transition metal FMs[17] and in AFM NiO/Pt bilayers excited via indirect heating[14]. The Q-fluence relation only shows slight deviations from linearity at the highest fluences. These deviations are potentially due to non-linearities in the absorption of the pump pulse.

### b) Time-resolved MLD measurements of $Mn_2Au$

Next, we turn to the metallic AFM $Mn_2Au$, in which the AFM order parameter is monitored via the purely electronic MLD effect described above. Figure 5a) shows the time-resolved MLD signals for absorbed fluences from ~0.95 mJ cm$^{-2}$ to ~9.55 mJ cm$^{-2}$.

The subsequent recovery occurs on a picosecond timescale, with some traces showing initially a subtle undershoot and overshoot pattern that may be related to optically induced strain waves. Fourier analysis does not reveal pronounced coherent oscillations attributable to magnons or phonons within the accessible signal-to-noise ratio. The maximum quenching Q again increases approximately linearly with absorbed fluence (see Fig. 5b).

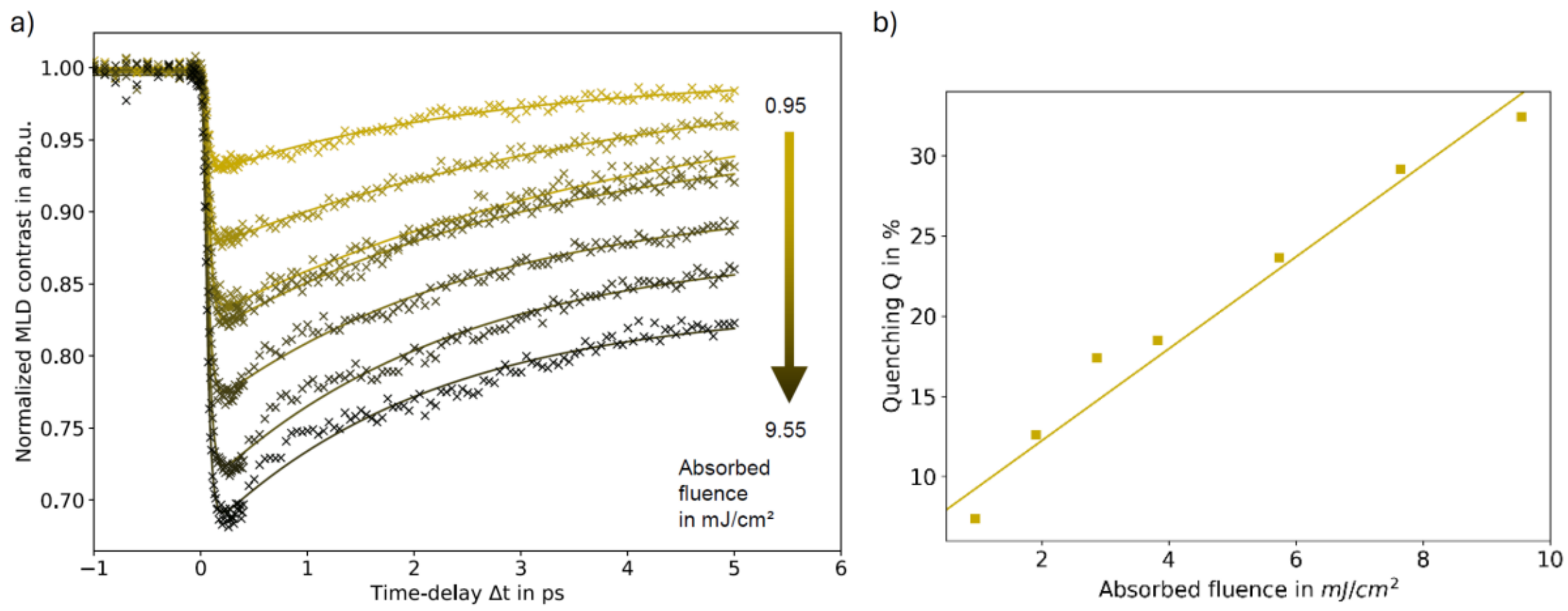

*Figure 5: a) Time-resolved MLD signal in Mn2Au for various absorbed pump fluences. Solid lines show double-exponential fits convolved with the temporal width of the pump pulse. b) Maximum quenching Q vs. absorbed fluence*

## Discussion

### Comparison between NiO, $Mn_2Au$, and the 3d ferromagnet Ni

Figure 6 summarizes the extracted demagnetization time $\tau_m$ of each magnetic sublattice as a function of the quenching (Q) for NiO and $Mn_2Au$. For comparison, we also show $\tau_m$ for a ferromagnetic Ni film. These Ni data were obtained from conventional time-resolved magneto-optical Kerr-effect measurements (see Supplemental Material). The thickness and substrate were chosen so that spin transport would not contribute to the measured magnetization dynamics of the Ni film. In Ni, $\tau_m$ increases approximately linearly with (Q), consistent with a three-temperature-model description incorporating Elliott–Yafet (EY)-type spin-flip scattering[5,18]. Within this framework, the slope of $\tau_m$ (Q), reflects the dependence of the demagnetization dynamics on the efficiency of energy transfer and spin-flip-mediated angular-momentum dissipation. By contrast, both NiO and $Mn_2Au$ exhibit an almost (Q)-independent $\tau_m$ over the investigated range. Beyond this qualitative difference in their quenching behavior, the three materials display markedly different characteristic timescales of $\tau_m$, highlighting the distinct roles of electronic structure and spin-flip probability in ultrafast magnetic-order dynamics.

These results address the first question raised in the Introduction: does the enhanced spin-flip probability $a_{sf}$ associated with magnetically compensated systems outweigh the suppression of scattering caused by an insulating gap? The comparison between the antiferromagnetic (AFM) charge-transfer insulator NiO and the metallic AFM $Mn_2Au$ provides a direct way to disentangle these competing effects. Despite the potentially different microscopic limitations on the extracted timescale $\tau_m$ in NiO, including the finite response time of lattice distortions coupled to the magnetic order, the observed difference in $\tau_m$ between NiO and $Mn_2Au$ remains substantial, amounting to approximately a factor of four. This contrast indicates that magnetic compensation alone does not determine the ultrafast response: the availability of electronic scattering channels is equally consequential. In NiO, the large charge-transfer gap strongly restricts the electronic phase space available for low-energy scattering. Consequently, conventional electron–electron scattering channels that can redistribute electronic energy and facilitate the reduction of magnetic order of each sublattices by joint antiparallel spin flips (see Fig. 1b) are suppressed relative to those in a metal. Although the AFM electronic structure can favor spin-flip processes through its compensated magnetic sublattice, an enhanced $a_{sf}$ cannot, by itself, compensate for a strongly reduced scattering phase space.

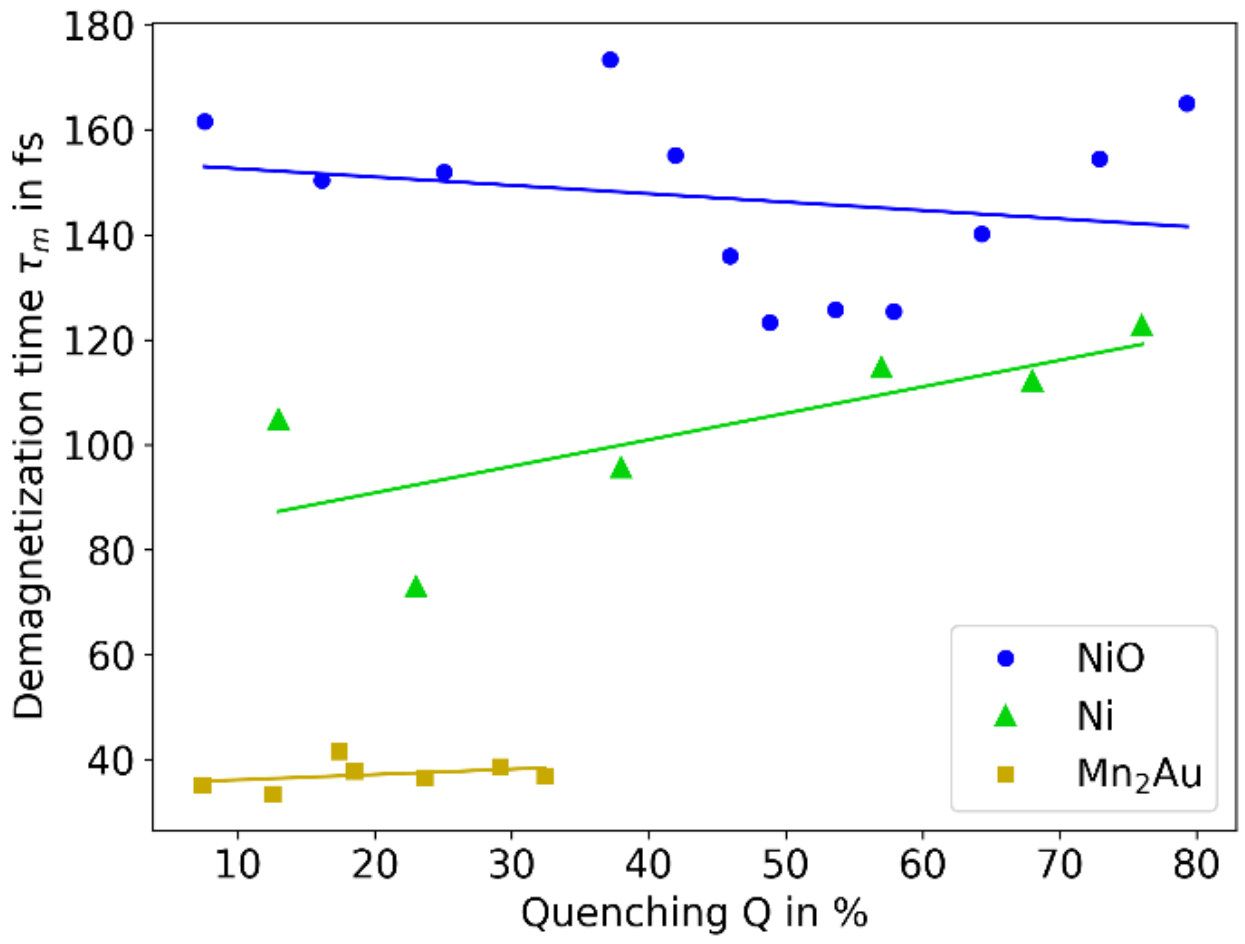

*Figure 6: Demagnetization time $\tau_m$ versus quenching Q for NiO (blue circles), $Mn_2Au$ (golden squares), and ferromagnetic Ni (green triangles). Linear fits are indicated as guide to the eye for each dataset*

The comparison with metallic ferromagnetic Ni supports this interpretation. The substantial density of electronic states around the Fermi level in Ni provides numerous available states for electron–electron scattering, including EY-type processes that couple spin relaxation to momentum scattering. Thus, despite the potentially larger spin-flip probability in the compensated AFM, the reduced electronic scattering phase space in insulating NiO leads to a longer demagnetization time $\tau_m$ than in metallic Ni. This comparison underscores that the ultrafast reduction of magnetic order is governed not solely by the probability of a spin-flip event, but also by the rate at which the underlying scattering events can occur.

The comparison between Ni and $Mn_2Au$ further addresses the second question raised in the Introduction: can the absence of a net magnetization and the associated relaxation constraints in a metallic AFM enable faster quenching of magnetic order than in a metallic ferromagnet? In contrast to NiO, metallic $Mn_2Au$ provides an appreciable electronic phase space for scattering of photoexcited hot electrons, while retaining the compensated AFM electronic structure. Under these conditions, the enhanced spin-flip probability $a_{sf}$ associated with the AFM sublattices can become a decisive factor in the ultrafast response. The resulting $\tau_m$ is substantially shorter than that of Ni, making $Mn_2Au$ one of the fastest-demagnetizing magnetic thin films reported (in the absence of spin transport). The comparison therefore suggests that the combination of metallic electronic phase space and AFM spin-dependent electronic structure, rather than magnetic compensation alone, can promote exceptionally rapid quenching of magnetic order.

Taken together, our results establish an important distinction between the *probability* of spin-flip scattering and the rate at which such scattering occurs. The former is influenced by the spin-dependent electronic structure and the availability of spin-mixing channels, whereas the latter depends on the electronic phase space and the relevant scattering matrix elements. A large $a_{sf}$ is therefore not sufficient to guarantee ultrafast magnetic-order dynamics if the available scattering channels are strongly restricted. Conversely, when these channels are activated by ultrashort pulse excitation, as in metallic $Mn_2Au$, the enhanced spin-flip probability $a_{sf}$ can translate into a particularly short $\tau_m$. This framework reconciles the contrasting responses of NiO, $Mn_2Au$, and Ni and identifies electronic phase space as a key factor governing the competition between insulating and metallic AFM dynamics.

Finally, it is important to distinguish the ultrafast quenching timescale discussed here from the characteristic speed commonly associated with AFM dynamics in the context of the Landau–Lifshitz–Gilbert (LLG) framework. As discussed in the Introduction, the latter generally refers to precessional motion and the characteristic frequencies of quasi-equilibrium magnetic excitations, whereas $\tau_m$ characterizes the nonequilibrium reduction of the magnetic-order parameter following optical excitation. These quantities describe physically distinct processes and need not exhibit the same material dependence. Our results thus do not establish a universal hierarchy of magnetic switching speeds; rather, they demonstrate that metallic AFMs can exhibit exceptionally rapid ultrafast quenching when their electronic structure supports efficient spin-flip-mediated relaxation. More broadly, they highlight the need to consider electronic phase space and spin-flip probability together when evaluating the potential of compensated magnetic materials for ultrafast spintronic applications.

## Conclusion

We have compared the laser-induced quenching of magnetic order in insulating NiO and metallic $Mn_2Au$ with the demagnetization of ferromagnetic Ni. Both antiferromagnets exhibit approximately quenching-independent demagnetization times over the investigated excitation range, whereas the demagnetization time in Ni increases with quenching. Despite their compensated magnetic order, the two antiferromagnets display markedly different responses: the extracted demagnetization time in $Mn_2Au$ is approximately four times shorter than in NiO and is also substantially shorter than in Ni. In insulating NiO, restricted electronic scattering channels limits the ultrafast response.

However, as the NiO measurement relies on a magnetically induced lattice distortion, its extracted timescale may represent an upper bound on the underlying magnetic response.

These findings support a picture in which rapid antiferromagnetic demagnetization requires both efficient spin-flip processes and sufficient electronic phase space for scattering. Magnetic compensation permits exchange-mediated scattering that reduces the order of both sublattices while conserving their combined spin angular momentum. However, this relaxation channel does not guarantee faster demagnetization when the underlying electronic scattering is restricted, as in a dielectric material. Metallic $Mn_2Au$ combines compensated magnetic order with accessible electronic scattering channels, providing favorable conditions for exceptionally rapid quenching. Although the comparison does not uniquely establish the microscopic scattering mechanism, it is consistent with this interplay between scattering rate and spin-flip probability.

## References

1. T. Kampfrath, A. Sell, G. Klatt, A. Pashkin, S. Mährlein, T. Dekorsy, M. Wolf, M. Fiebig, A. Leitenstorfer, and R. Huber, Coherent terahertz control of antiferromagnetic spin waves, *Nature Photonics* **5**, 31-34 (2011).
2. S. M. Rezende, A. Azevedo, and R. L. Rodríguez-Suárez, Introduction to antiferromagnetic magnons, *Journal of Applied Physics* **126**, 151101 (2019).
3. M. Arana, F. Estrada, D. S. Maior, J. B. S. Mendes, L. E. Fernandez-Outon, W. A. A. Macedo, V. M. T. S. Barthem, D. Givord, A. Azevedo, and S. M. Rezende, Observation of magnons in mn2au films by inelastic brillouin and raman light scattering, Applied Physics Letters **111**, 192409 (2017).
4. J. R. Sandercock and W. Wettling, Light scattering from surface and bulk thermal magnons in iron and nickel, *Journal of Applied Physics* **50**, 7784-7789 (1979).
5. B. Koopmans, J. J. M. Ruigrok, F. D. Longa, and W. J. M. de Jonge, Unifying ultrafast magnetization dynamics, *Physical Review Letters* **95**, 267207 (2005).
6. M. Krauß, T. Roth, S. Alebrand, D. Steil, M. Cinchetti, M. Aeschlimann, and H. C. Schneider, Ultrafast demagnetization of ferromagnetic transition metals: The role of the Coulomb interaction, *Physical Review B* **80**, 180407 (2009).
7. K. Leckron, A. Baral, and H.C. Schneider, Exchange scattering on ultrafast timescales in a ferromagnetic two-sublattice system, *Applied Physics Letters* **120**, 102407 (2022).
8. N. Thielemann-Kühn, D. Schick, N. Pontius, C. Trabant, R. Mitzner, K. Holldack, H. Zabel, A. Föhlisch, and C. Schüßler-Langeheine, Ultrafast and energy-efficient quenching of spin order: Antiferromagnetism beats ferromagnetism, *Physical Review Letters* **119**, 197202 (2017).
9. I. Radu, K. Vahaplar, C. Stamm, T. Kachel, N. Pontius, H.A. Dürr, T. A. Ostler, J. Barker, R.F. L. Evans, R. W. Chantrell, A. Tsukamoto, A. Itoh, A. Kirilyuk, Th. Rasing, and A. V. Kimel, Transient ferromagnetic-like state mediating ultrafast reversal of antiferromagnetically coupled spins, *Nature* **472**, 205-208 (2011)

10. W. L. Roth, Neutron and optical studies of domains in nio, *Journal of Applied Physics* **31**, 2000-2011 (1960).
11. C. Schmitt, L. Baldrati, L. Sanchez-Tejerina, F. Schreiber, A. Ross, M. Filianina, S. Ding, F. Fuhrmann, R. Ramos, F. Maccherozzi, D. Backes, M.-A. Mawass, F. Kronast, S. Valencia, E. Saitoh, G. Finocchio, and M. Kläui, Identification of néel vector orientation in antiferromagnetic domains switched by currents in nio/pt thin films, *Physical Review Applied* **15**, 034047 (2021).
12. H. J. Elmers, S. V. Chernov, S. W. D'Souza, S. P. Bommanaboyena, S. Y. Bodnar, K. Medjanik, S. Babenkov, O. Fedchenko, D. Vasilyev, S. Y. Agustsson, C. Schlueter, A. Gloskovskii, Y. Matveyev, V. N. Strocov, Y. Skourski, L. Šmejkal, J. Sinova, J. Minár, M. Kläui, G. Schönhense, and M. Jourdan, Néel vector induced manipulation of valence states in the collinear antiferromagnet mn2au, *ACS Nano* **14**, 17554 (2020).
13. J. Xu, C. Zhou, M. Jia, D. Shi, C. Liu, H. Chen, G. Chen, G. Zhang, Y. Liang, J. Li, W. Zhang, and Y. Wu, Imaging antiferromagnetic domains in nickel oxide thin films by optical birefringence effect, *Physical Review B* **100**, 134413 (2019).
14. S. Wust, C. Seibel, H. Meer, P. Herrgen, C. Schmitt, L. Baldrati, R. Ramos, T. Kikkawa, E. Saitoh, O. Gomonay, J. Sinova, Y. Mokrousov, H. C. Schneider, M. Kläui, B. Rethfeld, B. Stadtmüller, and M. Aeschlimann, Indirect optical manipulation of the antiferromagnetic order of insulating nio by ultrafast interfacial energy transfer, arXiv 2205.02686v2 (2022).
15. A. B. Shick, S. Khmelevskyi, O. N. Mryasov, J. Wunderlich, and T. Jungwirth, Spin-orbit coupling induced anisotropy effects in bimetallic antiferromagnets: A route towards antiferromagnetic spintronics, *Physical Review B* **81**, 212409 (2010).
16. V. Grigorev, M. Filianina, S. Y. Bodnar, S. Sobolev, N. Bhattacharjee, S. Bommanaboyena, Y. Lytvynenko, Y. Skourski, D. Fuchs, M. Kläui, M. Jourdan, and J. Demsar, Optical readout of the néel vector in the metallic antiferromagnet, *Physical Review Applied* **16**, 014037 (2021).
17. B. Koopmans, G. Malinowski, F. Dalla Longa, D. Steiauf, M. Fähnle, T. Roth, M. Cinchetti, and M. Aeschlimann, Explaining the paradoxical diversity of ultrafast laser-induced demagnetization, *Nature Mater* **9**, 259–265 (2010).
18. B. Koopmans, H. Kicken, M. van Kampen, and W. de Jonge, Microscopic model for femtosecond magnetization dynamics, *Journal of Magnetism and Magnetic Materials* **286**, 271-275 (2005).